\documentclass[a4paper]{jpconf}
\usepackage{graphicx}
\begin{document}
\title{Measurement of $\sigma_{\rm Total}$ in $e^{+}e^{-}$ Annihilations Below 10.56 GeV}

\author{Sheldon Stone}

\address{Physics Department, Syracuse University, Syracuse N.Y. 13244, USA}

\ead{stone@physics.syr.edu}

\begin{abstract}
Using the CLEO III detector, we measure absolute cross sections for
$e^+e^-\to$hadrons at seven center-of-mass energies between 6.964
and 10.538 GeV. $R$, the ratio of hadronic and muon pair production
cross sections, is measured at these energies with a r.m.s. error
$<$2\% allowing determinations of the strong coupling $\alpha_s$.
Using the expected evolution of $\alpha_s$ with energy we find
$\alpha_s(M_Z^2)=0.126\pm 0.005 ^{+0.015}_{-0.011}$, and
$\Lambda=0.31^{+0.09+0.29}_{-0.08-0.21}$.
\end{abstract}

\section{Introduction}
Theoretically $R(s)=\sigma_0(e^+e^-\to {\rm
hadrons})/\sigma_0(e^+e^-\to \mu^+\mu^-)$, where $s$ is the square
of the center-of-mass energy, provides a straight-forward way to
measure the strong coupling $\alpha_s$, since
$R(s)=R_0\left[1+C_1\frac{\alpha_s(s)}{\pi}+
C_2\left(\frac{\alpha_s(s)}{\pi}\right)^2
+C_3\left(\frac{\alpha_s(s)}{\pi}\right)^3+O(\alpha_s(s))^4\right]$,
where  $R_0$ is given by the number of color degrees of freedom (3)
times the sum of the squares of the quark charges. The $C_i$ are
determined by QCD calculations.

\section{Analysis Method}
The observed cross-section is the sum contributions from the bare
cross-section corrected by soft photon radiation (including virtual
higher order diagrams), $\sigma_{\rm sv}$,  a correction for hard
photon radiation, $\sigma_{\rm hard}$, and radiative tails from
resonant states, $\sigma_{\rm res}$. The ``Born" cross-section,
$\sigma_0=\sigma_{\rm sv}/\left(\epsilon(0)\delta_{\rm
sv}\right)$, where $\epsilon(0)$ is the efficiency for events
without initial state radiation and $\delta_{\rm sv}$ accounts for
soft photon emission and hadronic and leptonic vacuum polarization.
More details of the analysis method and the results are available
\cite{CLEO-paper}.

\section{Selection Criteria}
In order to suppress backgrounds from events other than $e^+e^- \to $~hadrons,
we apply selection requirements to individual tracks and showers
as well as to entire events. These cuts are not completely efficient and thus we need
to calculate their efficiencies from Monte Carlo simulation, thus leading to systematic errors, that dominate the uncertainties in our results.
Table~\ref{tab:trackandshower} lists the requirements for accepting
tracks and showers and individual events.

\begin{table}[htbp]
\caption{Requirements on Track \& Shower Selection, and Event
Selection.}  \label{tab:trackandshower}
\begin{center}
\begin{tabular}{ c l c l}
\hline\hline
\multicolumn{2}{c}{Track \& Shower}&\multicolumn{2}{c}{Event}\\
 Variable  & Allowed range&Variable  & Allowed range \\
\hline
$\chi^2$/NDF          &   $~~~<~100.0$ & $\mid Z_{\mbox{\scriptsize vertex}} \mid $ &   $ ~~~< 6.0$~cm \\
hit fraction          &   $~~~(0.5,1.2)$ & $E_{\mbox{\scriptsize vis}}/2E_{\mbox{\scriptsize beam}}$     &   $ ~~~>0.5 $\\
$\mid d_0 \mid$       &   $~~~<~3.0$~cm &$\mid P_{\mbox{\scriptsize z}}^{\mbox{\scriptsize miss}}/E_{\mbox{\scriptsize vis}} \mid $       &   $ ~~~< 0.3$\\
$\mid z_0 \mid $      &   $~~~<~18.0$~cm & $H_2/H_0   $               &   $~~~< 0.9 $\\
error of $z_0$        &   $~~~<~25.0$~cm & $E_{\mbox{\scriptsize cal}}/2E_{\mbox{\scriptsize beam}}$     &   $ ~~~(0.15,0.9) $\\
$\mid \cot(\theta) \mid$    &   $~~~<~3.0424$ & $E_{\gamma}^{\mbox{\scriptsize max}}/E_{\mbox{\scriptsize beam}}$   &   $~~~<0.8$\\
error of $\cot(\theta)$ &  $~~~<~0.50$ & $N_{\mbox{\scriptsize ChargedTrack}}$       &   $~~~\ge 4$ \\
 $P_{\rm track}/E_{\rm beam}$     &   $~~~(0.01,1.5)$  \\
 $E_{\rm shower}/E_{\rm beam}$      &   $~~~>0.01 $ \\
\hline\hline
\end{tabular}
\end{center}
\end{table}

\

Consistency with the beam collision point is enforced by the cut on
$d_0$, the distance of closest approach of the reconstructed track
relative to the beam axis, and on $z_0$, the distance between that
point and the average collision point on the beam axis.

\section{Results}
Besides estimating remaining backgrounds after these selection criteria are applied, we need to correct for beam radiation before annihilation interactions, that can then create
 $c\bar{c}$ and $b\bar{b}$ bound state resonances.
When the resonance decays to hadrons, our observed
cross section increases. For the purposes of our $R$
measurement, these contributions are sources of background and must also be
subtracted.

\begin{figure}[htbp]
\begin{center}
\includegraphics*[width=3.75in]{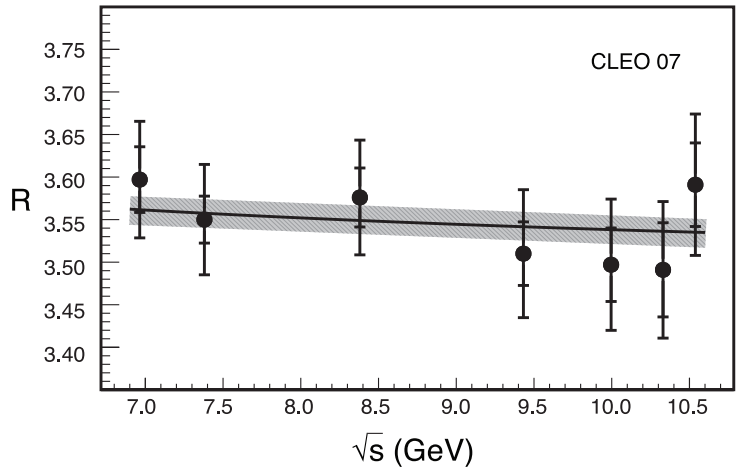}
\caption{CLEO $R$ measurements versus energy. The two sets of
uncertainties represent combined uncorrelated and statistical
uncertainties and total uncertainties; the line represents $R(s)$
with our average $\Lambda=0.31$~GeV, and the shaded area indicates
the $R$-values corresponding to one standard deviation in the
uncorrelated systematic uncertainty in $\Lambda$.}
\label{FigRvalues}
\end{center}
\end{figure}

The sources of systematic uncertainty for each continuum cross
section measurement include: luminosity, radiative correction,
trigger efficiency for hadronic events, multiplicity correction, and
hadronic event selection criteria. The resulting measured values of
$R$ as a function of energy are shown in Fig.~\ref{FigRvalues}.

Table~\ref{tab:alpha} shows the resulting $\alpha_s$ values obtained
at each of the seven energies. Comparing  $\alpha_s$ values with the
QCD predictions \cite{Bethke-07} at our energies, which assumes the
combined world average of $\alpha_s(M_{Z}^2) = 0.1189\pm0.0010$, we
find agreement within our quoted uncertainties.
\begin{table}[hbtp]
\caption{Measured values of $\alpha_s(s)$ with statistical and
systematic (common and uncorrelated) uncertainties, respectively.}
\label{tab:alpha}
\begin{center}
\begin{tabular}{cc}
\hline\hline
 $\sqrt s$ &  $\alpha_{s}(s)$\\
    GeV    &                 \\
\hline
~~~10.538 ~~~  &   ~~~$0.232~\pm~0.003~\pm~0.061~\pm~0.045$~~~ \\
   10.330      &  $0.142~\pm~0.005~\pm~0.051~\pm~0.049$ \\
    ~9.996     &  $0.147~\pm~0.004~\pm~0.057~\pm~0.038$ \\
    ~9.432     &  $0.159~\pm~0.004~\pm~0.058~\pm~0.033$ \\
    ~8.380     &  $0.218~\pm~0.022~\pm~0.053~\pm~0.023$ \\
    ~7.380     &  $0.195~\pm~0.017~\pm~0.052~\pm~0.018$ \\
    ~6.964     &  $0.237~\pm~0.030~\pm~0.052~\pm~0.018$ \\
\hline\hline
\end{tabular}
\end{center}
\end{table}

To test the compatibility with other measurements of $\alpha_s$ we use the
expected running of $\alpha_s$ with energy \cite{PDG-04}:
\begin{eqnarray}
\label{eq:FormR}
\alpha_s(s)=\frac{4\pi}{\beta_0\ln(s/\Lambda^2)}\left[ 1 -
\frac{2\beta_1}{\beta_0^2}\frac{\ln[\ln(s/\Lambda^2)]}{\ln(s/\Lambda^2)}
+\frac{4\beta_1^2}{\beta_0^4\ln^2(s/\Lambda^2)}\times\left(\left(\ln\left[\ln(s/\Lambda^2)\right]-\frac{1}{2}\right)^2
+\frac{\beta_2\beta_0}{8\beta_1^2}-\frac{5}{4}\right)\right] ,
\label{AlphaS}
\end{eqnarray}
where $n_f$ presents the number of quarks which have mass less than $\sqrt s$/2,
$\Lambda$ represents the QCD energy scale, and the $\beta$-functions are
defined as follows:
$\beta_0=11-2n_f/3$, $\beta_1=51-19n_f/3$, and $\beta_2=2857-5033n_f/9+325n_f^2/27$.

To find $\Lambda$, we use our $\alpha_s$ values at each energy point and
solve Eq.~(\ref{AlphaS}), assuming $n_f$ is equal to 4.
The value of $\Lambda$  varies from  0.11 at 10.330~GeV to 0.67
at 10.538~GeV. Using Eq.~(\ref{AlphaS}) with our average
value of  $\Lambda$, we extract the value of  the $\alpha_s$ at ${\sqrt s}=M_Z$.
Our results for $\alpha_s$ imply
$\Lambda = 0.31^{+0.09+0.29}_{-0.08-0.21}~\rm{GeV}$
and
$\alpha_s(M_{Z}^2) = 0.126\pm0.005~^{+0.015}_{-0.011}$,
where the uncertainties represent
statistical and total systematic, respectively.

Our results for $\alpha_s(M_{Z}^2)$ and $\Lambda(n_f=4)$ agree with the world averages
$\alpha_s(M_{Z}^2)=0.1189 \pm 0.0010$ \cite{Bethke-07} and
$\Lambda(n_f=4)=0.29 \pm 0.04$~GeV \cite{Bethke-04}.
Kuhn et al. \cite{Kuhn} (LTH 749) include quark mass effects and different matching between 4 and 5 flavor effective theories
They find using these data 
$\alpha_s(M_{Z}^2)=0.110^{+0.010+0.010}_{-0.012-0.011}$  and
$\Lambda=0.13^{+0.11+0.11}_{-0.07-0.07}$~GeV.

\section*{Acknowledgments}
This work was supported by the National Science Foundation. I thank
Surik Mehrabyan, Hector Mendez, and Karl Berkelman for useful
discussions.

\end{document}